# Observation of a phononic higher-order Weyl semimetal

Li Luo[1,#], Hai-Xiao Wang[2,3#], Bin Jiang[2], Ying Wu[1], Zhi-Kang Lin[2], Feng Li[1,†], Jian-Hua Jiang[2,†]

[1]School of Physics and Optoelectronics, South China University of Technology, 510640 Guangzhou, Guangdong, China

[2]School of Physical Science and Technology, and Collaborative Innovation Center of Suzhou Nano Science and Technology, Soochow University, 1 Shizi Street, Suzhou, 215006, China

[3]School of Physical Science and Technology, Guangxi Normal University, Guilin 541004, China

[#]These authors contributed equally to this work.

[†]Correspondence and requests for materials should be addressed to jianhuajiang@suda.edu.cn (Jian-Hua Jiang), jilinhubei@gmail.com (Feng Li).

## Abstract

**Weyl semimetals [1-4] are extraordinary systems where exotic phenomena such as Fermi arcs [5, 6], pseudo-gauge fields [7-9] and quantum anomalies [10-12] arise from topological band degeneracy in crystalline solids for electrons and metamaterials for photons [13-18] and phonons [19-22]. On the other hand, higher-order topological insulators [23-26] unveil intriguing multidimensional topological physics beyond the conventional bulk-edge correspondences. However, it is unclear whether higher-order topology can emerge in Weyl semimetals. Here, we report the experimental discovery of higher-order Weyl semimetals in its phononic analog which exhibit topologically-protected boundary states in multiple dimensions. We create the physical realization of the higher-order Weyl semimetal in a chiral phononic crystal with uniaxial screw symmetry. Using near-field spectroscopies, we observe the chiral Fermi arcs on the surfaces and a new type of hinge arc states on the hinge boundaries. These topological boundary arc states link the projections of Weyl points in different dimensions and directions, and hence demonstrate higher-order multidimensional topological physics in Weyl semimetals. Our study establishes the fundamental connection between higher-**

order topology and Weyl physics in crystalline materials and unveils a new horizon of higher-order topological semimetals where unprecedented materials such as higher-order topological nodal lines may emerge.

Symmetry plays a crucial role in the physics of topological materials. Symmetry-based indicators and topological quantum chemistry have been demonstrated as powerful theoretical tools for the diagnosis and prediction of topological insulators and semimetals [27-30]. Recent studies uncover symmetry-enforced topological Weyl and nodal-line semimetals in three-dimensional (3D) natural and artificial phononic and photonic crystals [22, 31-33]. In particular, artificial phononic crystals with remarkable controllability provide an excellent platform towards various topological phases protected by crystalline symmetries. Excellent examples include the lately discovered higher-order [34-40] and fragile [41] topological insulators in phononic crystals. However, the connection between the two fundamental classes of topological materials, the Weyl semimetals and higher-order topological insulators, though proposed very recently [42], is yet to be confirmed in experiments.

To fill this gap, here, we realize experimentally a novel topological phase: higher-order Weyl semimetals (HOWSMs) which exhibit simultaneously Weyl physics and higher-order topology. In conventional Weyl semimetals (WSMs), Weyl points (WPs) carry nontrivial topological chiral charge $N_C$ and the band topology is manifested as the two-dimensional (2D) chiral Fermi arcs that link the projections of the WPs with different chiral charges (see Figs. 1a and 1b). The chiral charge $N_C$, being equal to the change of the $k$-dependent Chern number (see Fig. 1c), indicates that WPs are the monopole sources or sinks of the Berry flux.

In HOWSMs, WPs can simultaneously carry the chiral charge $N_C$ and the higher-order charge $N_q$ (see Figs. 1d and 1e). The higher-order charge $N_q$, being equal to the change of the $k$-dependent higher-order topological numbers, reveals that WPs can be the pumping sources or annihilation sinks of the higher-order topology (see Fig. 1f and Supplementary Note 1). For instance, in Fig. 1d, such higher-order WPs (HOWPs) act as the transitions between the partial band gaps with finite Chern numbers and the partial band gap with a nontrivial quadrupole topological number. As a consequence, the HOWPs lead to one-dimensional (1D) hinge arc states linking the projections of the HOWPs with opposite $N_q$, in addition to the 2D chiral Fermi arcs (Fig. 1e). The coexisting chiral and higher-order charges of HOWPs in HOWSMs lead to rich physics, as illustrated below. Meanwhile, in the bulk, the extraordinary properties of WPs, e.g., pseudo gauge fields and chiral anomaly, remain intact for HOWPs, since these properties originate from the chiral charges.

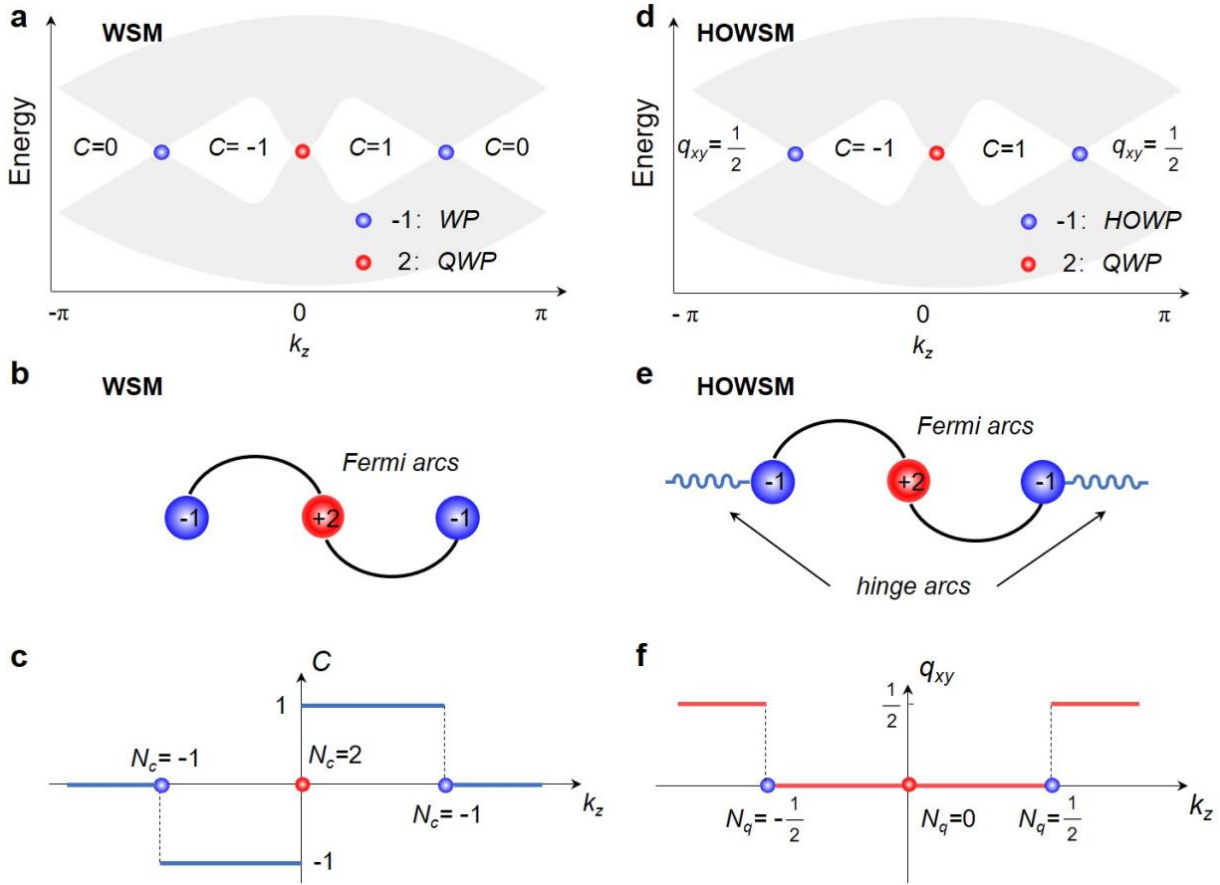

**Figure 1 | Higher-order Weyl semimetal**. **a-c**, An example of conventional Weyl semimetals (WSMs) with two Weyl points (WPs) and a quadratic Weyl point (QWP). **a,** Schematic of the bulk bands and the partial band gaps with $k_z$-dependent Chern numbers $C$. **b,** Schematic of chiral Fermi arcs linking the QWP and WPs. **c**, The $k_z$-dependent Chern numbers $C$ and the chiral charge of the QWP and WPs $N_C$. **d-f**, An example of higher-order Weyl semimetals (HOWSMs) with two higher-order WPs (HOWPs) and a QWP. **d,** Schematic of the bulk bands and the partial band gaps with $k_z$-dependent Chern numbers $C$ and quadrupole topological numbers $q_{xy}$ in the HOWSM. **e,** Schematic of the coexisting 2D Fermi arcs and 1D hinge arcs and their connectivity with the HOWPs. **f,** The $k_z$-dependent higher-order topological numbers $q_{xy}$ and the higher-order charges of the WPs $N_q$.

We design an air-borne phononic crystal to realize the HOWSM in its phononic analog. The phononic crystal, fabricated via 3D-printing technology using photosensitive resin (Figs. 2a and 2b), forms a tetragonal lattice with lattice constants $a = 20$ mm and $a_z =28$ mm for the $x$-$y$ plane and the $z$ direction, respectively. Other geometry parameters are specified in the Supplementary Note 1.

In our design, both the Weyl physics and the higher-order topology are induced by the chiral crystalline symmetry (space group $P4_122$, no. 91). Particularly, the screw symmetry (Fig. 2c), $S_{4z}: (x, y, z) \rightarrow (y, -x, z + \frac{a_z}{4})$, plays an essential role. The screw symmetry, together with other crystalline symmetries, quantize the quadrupole topological number $q_{xy}$ and lead to the higher-order topology at large $|k_z|$. Meanwhile, the screw symmetry also leads to synthetic gauge flux and finite Chern numbers, $C = \pm 1$, at small $|k_z|$ (see Supplementary Note 1). We denote the small $|k_z|$ region with finite Chern numbers as the Chern partial gaps (CPGs), while the large $|k_z|$ region with a nontrivial quadrupole topological number $q_{xy} = \frac{1}{2}$ as the higher-order partial gap (HOPG). The HOWPs at $\bm{k} = (\pi, \pi, \pm k_{WP})$ lie at the boundaries between the CPGs and the HOPG (Figs. 2d-2f). In addition, there is a quadratic WP (QWP) with a chiral charge $N_C = 2$ at the Brillouin zone center which separates the CPGs with opposite Chern numbers (Figs. 2d-2f). Figure 2d shows that the phononic crystal has a clean spectrum which are perfect for the study of the properties of HOWSMs.

We use near-field spectroscopies to study the spectral properties at the surfaces and hinges. The experimental setup for the surface near-field spectroscopy is depicted in Fig. 3a. A subwavelength acoustic source (S$_1$) is placed at the center of the YZ surface where a resin plate is fabricated to form a hard-wall surface boundary. A small hole is open at the center of the surface boundary to insert the acoustic source. A tiny microphone is inserted into the phononic crystal to probe the surface acoustic fields right underneath the hard-wall boundary. By scanning the acoustic field distributions underneath the entire surface boundary for various frequencies and then Fourier transforming the detected acoustic fields, we can extract the dispersions of the surface acoustic states.

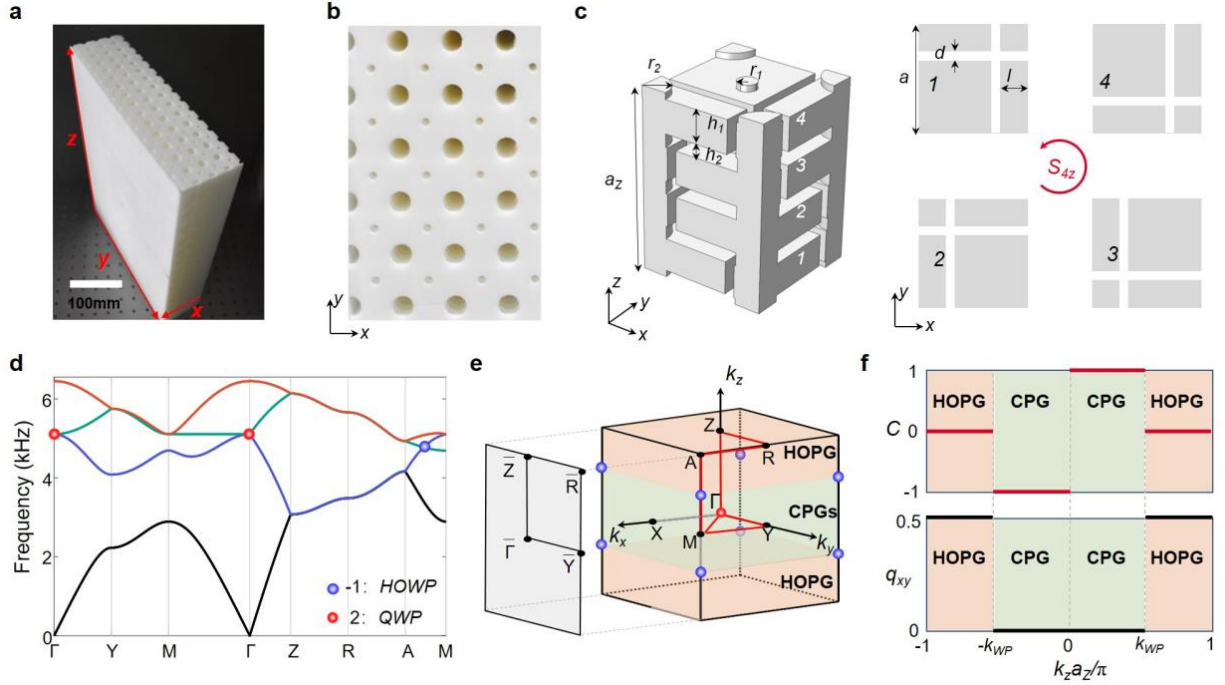

**Figure 2 | Phononic higher-order Weyl semimetal**. **a-b**, Photographs of the fabricated phononic crystal. **c**, Left: Unit-cell structure of the phononic crystal. Gray solid regions represent the air regions where the acoustic waves propagate. The unit-cell consists of four layers connected vertically via the cylindrical holes at the cell center and the cell hinges with radii $r_1$ and $r_2$, respectively. The geometry parameters are specified in Supplementary Note 1. These four layers can be transformed into one another through the $S_{4z}$ symmetry as illustrated in the right panel. **d**, Bulk phononic band structure of the HOWSM. The QWP and HOWPs are depicted. **e**, The bulk and surface Brillouin zones. The Chern partial gaps (CPGs) and higher-order partial gaps (HOPGs) are illustrated. **f**, The $k_z$-dependent Chern and higher-order topological numbers in the phononic HOWSM.

Figure 3b shows the detected acoustic field distribution right below the YZ surface when the source S1 has a frequency of 4.7kHz. The highly directional acoustic field pattern, which implies the openness of the iso-frequency contour of the surface acoustic waves, is a direct indication of the Fermi arc surface states [20]. The measured dispersions of the surface acoustic waves for various $k_z$ are presented in Figs. 3c-3f which show excellent agreement with the simulation. In Fig. 3c, with $k_z = 0$, the measured phononic dispersions confirm the existence of the QWP at the Γ point (red sphere). In Fig. 3d, with $k_z = \frac{0.3\pi}{a_z}$, the CPG with Chern number $C = 1$ is visualized experimentally through the gapless chiral edge states. In Fig. 3e, with $k_z = k_{WP} = \frac{0.55\pi}{a_z}$, the HOWP is found as the gap closing point at $k_y = \frac{\pi}{a}$ (blue spheres). Although the bulk band gap closes, the edge states still appear and merge into the HOWP, which can be understood as the residue effect of the helicoid Fermi arc surface states [43]. In Fig. 3f, at $k_z = \frac{\pi}{a_z}$, the edge states become gapped, a feature which is consistent with the higher-order topology at $|k_z| > k_{WP}$. From these measured dispersions, the transitions from the CPG to HOPG are clearly witnessed where the HOWP serves as the transition point between them.

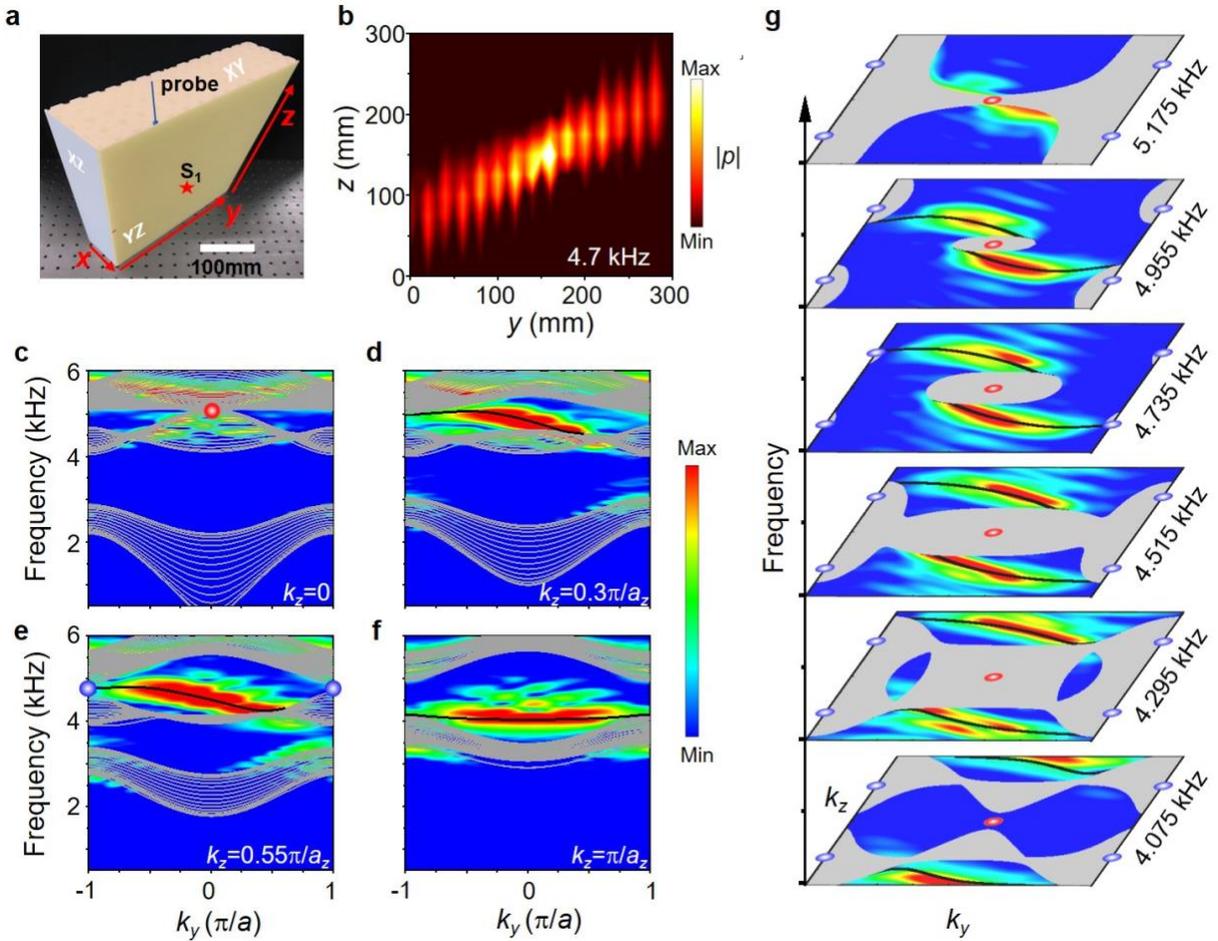

**Figure 3 | Topological surface states of the HOWSM**. **a**, Illustration of the experimental setup for the near-field spectroscopy. **b**, Measured amplitude of the acoustic pressure $|p|$ right below the YZ surface as excited by the source $S_1$ at the center of the YZ surface. **c-f**, Measured dispersions of the surface waves (color) with $k_z = 0$, $0.3\pi/a_z$, $0.55\pi/a_z$ and $\pi/a_z$, separately. Gray curves are the

calculated bulk dispersions. Black curves are the calculated dispersions of the YZ surface states. Red and blue spheres denote the projections of the QWP and HOWPs, respectively. **g**, Measured iso-frequency contours (color) of the YZ surface waves for various frequencies. Gray regions represent the calculated bulk bands. Black curves denote the calculated iso-frequency contours of the YZ surface states.

To visualize directly the Fermi arc surface states, we present the iso-frequency contours of the surface acoustic waves in Fig. 3g at various frequencies. From 5.175kHz to 4.515kHz, the emergence of the long chiral Fermi arcs linking the projections of the QWP and the HOWPs are clearly visible. Such very long chiral Fermi arcs ($\sim \frac{\pi}{a}$) are developed due to the highly chiral structure of the phononic crystal. The winding of the chiral Fermi arcs around the QWP is consistent with the picture of helicoid surface states in WSMs [43-45]. In addition to the chiral Fermi arcs, gapped surface states due to higher-order topology emerge at $|k_z| > k_{WP}$, as shown in Fig. 3f. The iso-frequency contour at 4.075kHz indicates such gapped surface states that do not connect the projections of the QWP and the HOWPs. An intermediate case with frequency 4.075kHz shows the crossover between the chiral Fermi arc states and the gapped surface states.

The topological hinge states are measured using a similar near-field scanning method. Before measuring the hinge states, it is necessary to show that the hinge states are solely due to the higher-order topology. In our phononic crystal, if the inter-cell and intra-cell couplings are switched (by interchanging the geometry parameters $r_1$ and $r_2$), the higher-order topology and the hinge states can be eliminated. Although the bulk phonon dispersions remain the same, such alteration leads to a conventional Weyl semimetal (see Supplementary Note 2 for details). We then verify numerically the bulk-hinge correspondence by simulating the acoustic wave propagation when a source is placed at the top-end of the hinge (purple stars in Fig. 4a, S$_2$) with frequency 4.033kHz. For the conventional WSM, the excited acoustic wave spreads into the surfaces and are not confined on the hinge. In contrast, for the HOWSM, the acoustic wave excited by the same source is well-confined by and propagates along the hinge boundary. The emergence and propagation of the hinge states in the HOWSM is also confirmed in experiments, as shown in Fig. 4b.

We then measure the dispersions of the hinge states using the near-field spectroscopy. By setting the source at the bottom-end of the hinge, we can excite the hinge states with positive group velocities, i.e., the hinge states propagating upwards. Figure 4c shows the measured spectrum of such hinge states which is consistent with the calculation. By placing the acoustic source at the top-end of the hinge, we can excite the downward-propagating hinge states. The measured spectrum of such hinge states with negative group velocities also agrees with the calculation. The observed hinge arcs are terminated at the projections of the HOWPs, which confirms the picture illustrated in Fig. 1e.

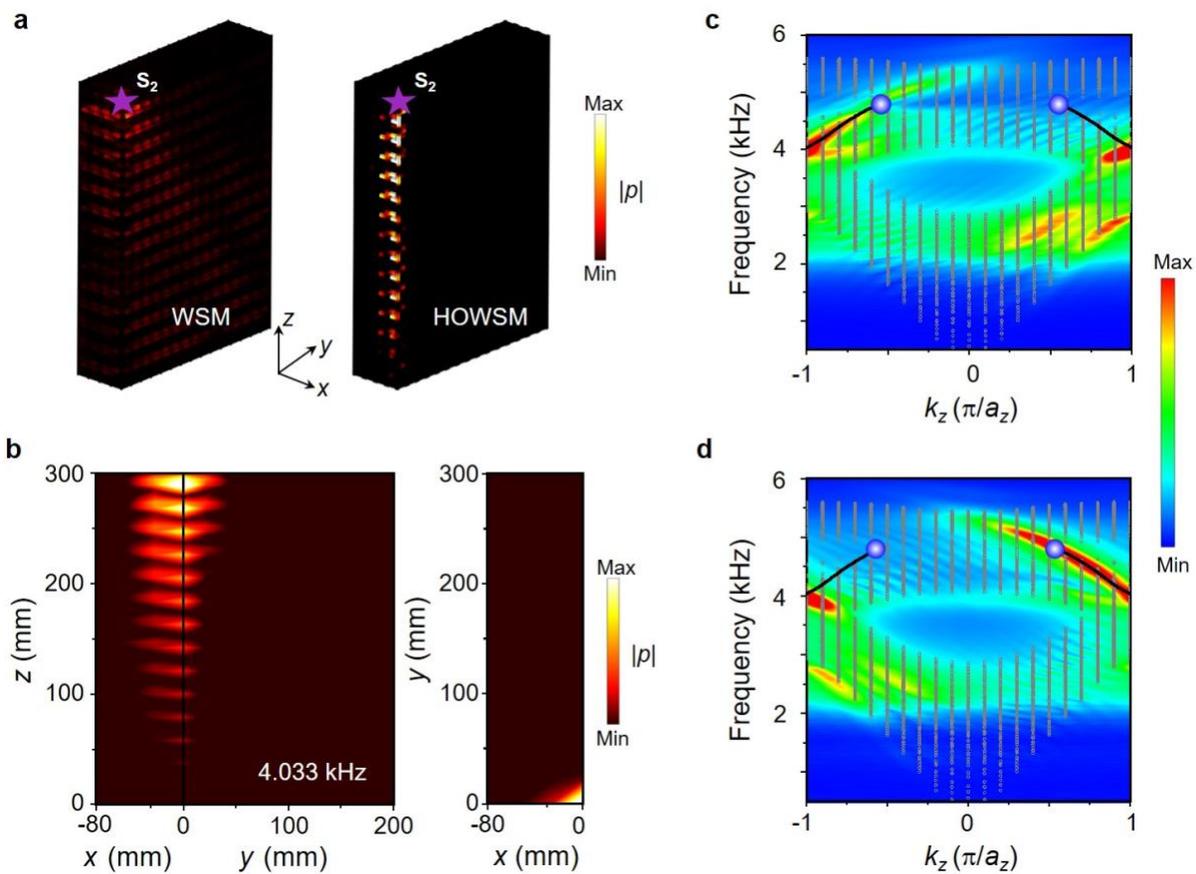

**Figure 4 | Higher-order hinge arc states**. **a**, Simulated wave propagations in the phononic WSM and HOWSM when an acoustic source with frequency 4.033kHz is placed at the top-end of the hinge boundary (purple stars, $S_2$). **b**, Measured acoustic wave propagation along the hinge for the HOWSM. **c-d**, Measured dispersions of the hinge states (color) when the acoustic source is placed, respectively, at the bottom- and top-end of the hinge boundary. Black curves represent the calculated dispersions of

the hinge states. Gray dots represent the calculated edge and bulk states. Blue spheres denote the projections of the HOWPs.

In conclusion, we experimentally discovered a new topological phase, the HOWSMs, which exhibit unique multidimensional topological properties. The ability of integrating the Weyl points in the 3D bulk, the chiral Fermi arc states on the 2D surfaces and the hinge arc states on the 1D hinges in HOWSMs may yield interesting applications such as topologically-protected integrated phononic devices. Realizing HOWSMs in electronic systems may lead to materials with fractional electronic charges at the hinge boundaries [42] and other exotic properties. Furthermore, this work opens a new research frontier of higher-order topological semimetals where many novel topological phases such as higher-order topological nodal lines and higher-order multifold Weyl/Dirac points are yet to be discovered.

## Materials and Methods

### Simulations

All simulations in this paper are implemented with the acoustics module of COMSOL Multiphysics. The speed of sound and the air density used are 343m/s and 1.29kg/m$^3$, respectively. To obtain the bulk bands (Fig. 2d), boundaries of the unit cell in three directions are set to be periodic. In order to calculate the dispersions of the surface wave (Fig. 3), the hard boundary conditions in $x$-direction and periodic boundary conditions in the remaining directions are applied to the supercell composed of 15 unit-cells along the $x$-direction. The surface wave dispersions are obtained by scanning the wavevectors in the entire surface Brillouin zone. To simulate the acoustic field distributions at the hinge, we calculate the frequency response of conventional and higher-order Weyl semimetals with 4, 15, 15 periods in $x$-, $y$- and $z$-directions, respectively (Fig. 4). The acoustic sources (purple stars in Fig. 4a) are set at the top end of the hinge. To obtain the dispersions of the hinge states, a supercell with 11, 11, 1 periods along the $x$-, $y$-, and $z$-directions, respectively, is constructed. Moreover, periodic boundary and the hard-wall boundary conditions are imposed in the $z$ direction and the two side surfaces, respectively. The remaining side surfaces are set to radiation boundary conditions.

## Experiments

The sample with 4, 15, 15 periods in the *x*-, *y*- and *z*-directions, respectively, is manufactured by 3D printing technology using photosensitive resin. The 2 mm thickness boundaries cover two sides of the sample, while other sides are kept open. To measure the surface state, a headphone (diameter 6mm) used for sound excitation is placed in the cavity near YZ surface of the sample (Fig. 3a, denoted as $S_1$). Afterwards, a tiny microphone is inserted into the sample to measure the acoustic fields underneath the YZ surface using the network analyzer (Keysight E5061B). The measurement is achieved by the microphone scanning along the *y*- and *z*-directions with scanning steps of 20 mm and 14 mm, respectively. By implementing the 2D Fourier transformations of the measured acoustic fields, the surface dispersion and surface Fermi arcs (Fig. 3) are obtained. For the measurements of the hinge states, the acoustic source is placed at the top- or bottom-end of the hinge, then we put the probe into the sample and measured the acoustic fields by scanning the entire sample. The dispersions of the hinge states (Fig. 4) are unfolded by performing the 1D Fourier transformation for the measured acoustic pressure fields along the hinge.

## Acknowledgements

H.X.W, B.J, Z.K.L and J.H.J are supported by the National Natural Science Foundation of China (Grant No. 12074281) and the Jiangsu Province Specially-Appointed Professor Funding. L.L, Y.W and F.L are supported by the Natural Science Foundation of Guangdong Province (No. 2020A1515010549), China Postdoctoral Science Foundation (NO. 2020M672615).

## Author contributions

J.H.J initiated the project. J.H.J and F.L guided the research. J.H.J, H.X.W, B.J and Z.K.L established the theory. H.X.W and L.L performed the numerical calculations and simulations. L.L, Y.W, J.H.J and F.L designed and achieved the experimental set-up and the measurements.

All the authors contributed to the discussions of the results and the manuscript preparation. J.H.J, H.X.W and F.L wrote the manuscript and the Supplementary Information.

## Competing Interests

The authors declare that they have no competing financial interests.

## Data availability

All data are available in the manuscript and the Supplementary Information. Additional information is available from the corresponding authors through proper request.

## Code availability

We use the commercial software COMSOL MULTIPHYSICS to perform the acoustic wave simulations and eigenstates calculations. Request to computation details can be addressed to the corresponding authors.

## References

[1] Wan, X., Turner, A. M., Vishwanath, A. & Savrasov, S. Y. Topological semimetal and Fermi-arc surface states in the electronic structure of pyrochlore iridates. *Phys. Rev. B* **83**, 205101 (2011).

[2] Xu, G., Weng, H., Wang, Z., Dai, X. & Fang, Z. Chern semimetal and the quantized anomalous Hall effect in $HgCr_2Se_4$. *Phys. Rev. Lett.* **107**, 186806 (2011).

[3] Burkov, A. A. & Balents, L. Weyl semimetal in a topological insulator multilayer. *Phys. Rev. Lett.* **107**, 127205 (2011).

[4] Armitage, N. P., Mele, E. J. & Vishwanath, A. Weyl and Dirac semimetals in three-dimensional solids. *Rev. Mod. Phys.* **90**, 015001 (2018).

[5] Xu, S.-Y. *et al.* Discovery of a Weyl fermion semimetal and topological Fermi arcs. *Science* **349**, 613-617 (2015).

[6] Lv, B. Q. *et al.* Experimental discovery of Weyl semimetal TaAs. *Phys. Rev. X* **5**, 031013 (2015).

[7] Jia, H. *et al.* Observation of chiral zero mode in inhomogeneous three-dimensional Weyl metamaterials. *Science* **363**, 148-151 (2019).

[8] Peri, V., Serra-Garcia, M., Ilan, R. & Huber, S. D. Axial-field-induced chiral channels in an acoustic Weyl system. *Nat. Phys.* **15**, 357-361(2019).

[9] Ilan, R., Grushin, A. G. & Pikulin, D. I. Pseudo-electromagnetic fields in 3D topological semimetals. *Nat. Rev. Phys.* **2**, 29-41 (2020).

[10] Huang, X. *et al.* Observation of the chiral-anomaly-induced negative magnetoresistance in 3D Weyl semimetal TaAs. *Phys. Rev. X* **5**, 031023 (2015).

[11] Hirschberger, M., Kushwaha, S., Wang, Z., Gibson, Q., Liang, S., Belvin, C. A., Bernevig, B. A., Cava R. J. & Ong, N. P. The chiral anomaly and thermopower of Weyl fermions in the half-Heusler GdPtBi. *Nat. Mater.* **15**, 1161-1165 (2016).

[12] Gooth, J. *et al.* Experimental signatures of the mixed axial–gravitational anomaly in the Weyl semimetal NbP. *Nature* **547**, 324-327 (2017).

[13] Lu, L., Fu, L., Joannopoulos, J. D. & Soljačić, M. Weyl points and line nodes in gyroid photonic crystals. *Nat. Photonics* **7**, 294-299 (2013).

[14] Lu, L., Wang, Z., Ye, D., Ran, L., Fu, L., Joannopoulos, J. D. & Soljačić, M. Experimental observation of Weyl points. *Science* **349**, 622-624 (2015).

[15] Chen, W. J., Xiao, M. & Chan, C. T. Photonic crystals possessing multiple Weyl points and the experimental observation of robust surface states. *Nat. Commun.* **7**, 13038 (2016).

[16] Noh, J., Huang, S., Leykam, D., Chong, Y. D., Chen, K. P. & Rechtsman, M. C. Experimental observation of optical Weyl points and Fermi arc-like surface states. *Nat. Phys.* **13**, 611 (2017).

[17] Yang, B. *et al.* Ideal Weyl points and helicoid surface states in artificial photonic crystal structures. *Science* **359**, 1013-1016 (2018).

[18] Wang, D. *et al.* Photonic Weyl points due to broken time-reversal symmetry in magnetized semiconductor. *Nat. Phys.* **15**, 1150-1155 (2019).

[19] Yang, Z. & Zhang, B. Acoustic type-II Weyl nodes from stacking dimerized chains. *Phys. Rev. Lett.* **117**, 224301 (2016).

[20] Li, F., Huang, X., Lu, J., Ma, J. & Liu, Z. Weyl points and Fermi arcs in a chiral phononic crystal. *Nat. Phys.* **14**, 30-34 (2018).

[21] He, H., Qiu, C., Ye, L., Cai, X., Fan, X., Ke, M., Zhang, F. & Liu, Z. Topological negative refraction of surface acoustic waves in a Weyl phononic crystal. *Nature* **560**, 61-64 (2018).

[22] Yang, Y. *et al.* Topological triply degenerate point with double Fermi arcs. *Nat. Phys.* **15**, 645-649 (2019).

[23] Benalcazar, W. A., Bernevig, B. A. & Hughes, T. L. Quantized electric multipole insulators. *Science* **357**, 61-66 (2017).

[24] Langbehn, J., Peng, Y., Trifunovic, L., von Oppen, F. & Brouwer, P. W. Reflection-symmetric second-order topological insulators and superconductors. *Phys. Rev. Lett.* **119**, 246401 (2017).

[25] Song, Z. D., Fang, Z. & Fang, C. (d-2)-dimensional edge states of rotation symmetry protected topological states. *Phys. Rev. Lett.* **119**, 246402 (2017).

[26] Schindler, F. *et al.* Higher-order topological insulators. *Sci. Adv.* **4**, eaat0346 (2018).

[27] Bradlyn, B., Elcoro, L., Cano, J., Vergniory, M. G., Wang, Z., Felser, C., Aroyo, M. I. & Bernevig, B. A. Topological quantum chemistry. *Nature* **547**, 298-305(2017).

[28] Vergniory, M. G., Elcoro, L., Felser, C., Regnault, N., Bernevig, B. A. & Wang, Z. A complete catalogue of high-quality topological materials. *Nature* **566**, 480-485 (2019).

[29] Tang, F., Po, H. C., Vishwanath, A. & Wan, X. Comprehensive search for topological materials using symmetry indicators. *Nature* **566**, 486-489 (2019).

[30] Zhang, T., Jiang, Y., Song, Z., Huang, H., He, Y., Zhong, F., Weng, H. & Fang, C. Catalogue of topological electronic materials. *Nature* **566**, 475-479 (2019).

[31] Zhang, T., Song, Z., Alexandradinata, A., Weng, H., Fang, C., Lu, L. & Fang, Z. Double-Weyl phonons in transition-metal monosilicides. *Phys. Rev. Lett.* **120**, 016401 (2018).

[32] Xiong, Z., Wang, H.-X., Ge, H., Shi, J., Luo, J., Lai, Y., Lu, M.-H. & Jiang, J.-H. Topological node lines in mechanical metacrystals. *Phys. Rev. B* **97**, 180101 (2018).

[33] Liu, Q.-B., Qian, Y., H.-H. Fu & Wang, Z. Symmetry-enforced Weyl phonons. Preprint at https://www.arXiv.org/abs/1911.07461

[34] Serra-Garcia, M. *et al.* Observation of a phononic quadrupole topological insulator. *Nature* **555**, 342-345 (2018).

[35] Peterson, C. W., Benalcazar, W. A., Hughes, T. L. & Bahl, G. A quantized microwave quadrupole insulator with topological protected corner states. *Nature* **555**, 346-350 (2018).

[36] Imhof, S. *et al.* Topolectrical circuit realization of topological corner modes. *Nat. Phys.* **14**, 925-929 (2018).

[37] Noh, J. *et al.* Topological protection of photonic mid-gap defect modes. *Nat. Photon.* **12**, 408-415 (2018).

[38] Xue, H., Yang, Y., Gao, F., Chong, Y. & Zhang, B. Acoustic higher-order topological insulator on a kagome lattice. *Nat. Mater.* **18**, 108-112 (2019).

[39] Ni, X., Weiner, M., Alù, A. & Khanikaev, A. B. Observation of higher-order topological acoustic states protected by generalized chiral symmetry. *Nat. Mater.* **18**, 113-120 (2019).

[40] Zhang, X. *et al.* Second-order topology and multi-dimensional topological transitions in sonic crystals. *Nat. Phys.* **15**, 582-588 (2019).

[41] Peri, V. *et al.* Experimental characterization of fragile topology in an acoustic metamaterial. Science 367, 797-800 (2020).

[42] Wang, H.-X., Lin, Z.-K., Jiang, B., Guo, G.-Y. & Jiang, J.-H. Higher-order Weyl semimetals. *Phys. Rev. Lett.* **125**, 146401 (2020).

[43] Fang, C., Lu, L., Liu, J. & Fu, L. Topological semimetals with helicoid surface states. *Nat. Phys.* **12**, 936-941 (2016).

[44] Rao, Z. *et al.* Observation of unconventional chiral fermions with long Fermi arcs in CoSi. *Nature* **567**, 496-499 (2019).

[45] Sanchez, D. S. *et al.* Topological chiral crystals with helicoid-arc quantum states. *Nature* **567**, 500-505 (2019).